# Light and microwave driven spin pumping across FeGaB – BiSb interface

Vinay Sharma[1], Weipeng Wu[2], Prabesh Bajracharya[1], Duy Quang To[3, 2], Anthony Johnson[1], Anderson Janotti[3], Garnett W. Bryant[4], Lars Gundlach[2, 5], M. Benjamin Jungfleisch[2, *] and Ramesh C. Budhani[1, **]

[1]*Department of physics, Morgan State University, Baltimore, 21251 MD USA*

[2]*Department of Physics and Astronomy, University of Delaware, Newark, DE 19716, USA*

[3]*Department of Materials Science and Engineering, University of Delaware, Newark, DE 19716, USA*

[4]*Nanoscale Device Characterization Division and Joint Quantum Institute, National Institute of Standards and Technology, 100 Bureau Drive, Gaithersburg, MD 20899, USA*

[5]*Department of Chemistry and Biochemistry, University of Delaware, Newark, DE 19716, USA*

[*]*mbj@udel.edu*

[**]*ramesh.budhani@morgan.edu*

**Abstract**

3-D topological insulators (TI) with large spin Hall conductivity have emerged as potential candidates for spintronic applications. Here, we report spin to charge conversion in bilayers of amorphous ferromagnet (FM) $Fe_{78}Ga_{13}B_9$ (FeGaB) and 3-D TI $Bi_{85}Sb_{15}$ (BiSb) activated by two complementary techniques: spin pumping and ultrafast spin-current injection. DC magnetization measurements establish the soft magnetic character of FeGaB films, which remains unaltered in the heterostructures of FeGaB-BiSb. Broadband ferromagnetic resonance (FMR) studies reveal enhanced damping of precessing magnetization and large value of spin mixing conductance ($5.03 \times 10^{19}$ $m^{-2}$) as the spin angular momentum leaks into the TI layer. Magnetic field controlled bipolar dc voltage generated across the TI layer by inverse spin Hall effect is analyzed to extract the values of spin Hall angle and spin diffusion length of BiSb. The spin pumping parameters derived from the measurements of the femtosecond light-pulse-induced terahertz emission are consistent with the result of FMR. Kubo-Bastin formula and tight-binding model calculations shed light on the thickness-dependent spin-Hall conductivity of the TI films, with predictions that are in remarkable agreement with the experimental data. Our results suggest that room temperature deposited amorphous and polycrystalline heterostructures provide a promising platform for creating novel spin orbit torque devices.

## I. Introduction

The exchange of spin angular momentum across the interface between a magnetically ordered material and a non-magnetic heavy metal with strong spin-orbit coupling (SOC) or a semimetal of non-trivial band topology has been of considerable fundamental and technological interest in recent years [1, 2]. The fact that this exchange can be initiated by excitations of varied frequencies ranging from direct current to optical radiation, and its manifestations as phenomena of various time scales has opened opportunities for creating new breeds of spintronic devices. This potential for applications is also the motivator of search for new material combinations that show stronger interface effects, preferably at ambient temperature. Topological insulators (TIs) have emerged as a new class of materials for efficient conversion of spin to charge current and vice-versa [3, 4]. This efficiency of TIs has been evaluated by measuring the microwave induced inverse spin-Hall effect (ISHE), spin-torque ferromagnetic resonance (ST-FMR) [5, 6], thermally induced spin injection [7] and through the measurement of non-linear Hall resistance [8]. Most of the FM-TI heterostructures based spin-pumping experiments reported in the literature are performed on crystalline ferromagnetic metals (FM) and ferrimagnetic insulators interfaced with epitaxial films of the topological insulator $Bi_2Se_3$ [3, 9, 10]. A rather simple alternative to $Bi_2Se_3$ is offered by the binary alloy $Bi_{1-x}Sb_x$, which for values of x in the interval 0.03 to 0.22 is a three-dimensional TI [11, 12] as established by angle resolved photoemission spectroscopy on single crystals [13] and epitaxial thin films [14]. Electronic transport measurements on such crystals are characterized by a metal-like resistivity at low temperatures and the presence of weak Shubnikov de Haas oscillations in the longitudinal ($\rho_{xx}$) resistivity [12]. The topological phase of $Bi_{1-x}Sb_x$ has been identified as an excellent spin-orbit-torque (SOT) material as revealed by dc transport measurements on epitaxial bilayers of $Bi_{1-x}Sb_x$ and a ferromagnet like MnGa [8]. Interestingly, two recent dc transport studies [15, 16] have also indicated that polycrystalline films of $Bi_{1-x}Sb_x$ made by a scalable process like sputtering are quite effective in producing spin currents to drive the magnetization of FeMn, FeCoB and CoTb thin films by SOT. This remarkable result invites detailed investigations of spin pumping across a clean $Bi_{1-x}Sb_x$ – FM interface with techniques of various strength and time scales. A clear interpretation of the spin pumping and THz emission results is also easier if the FM counterpart of the interface is made of a soft ferromagnet with in-plane magnetic anisotropy and no preferred direction of magnetization. This requirement is met by films of $Fe_{78}Ga_{13}B_9$ (FeGaB) metallic glass, which is known for its soft magnetic character.

In this work we address the spin-to-charge conversion process in FeGaB ferromagnet interfaced with $Bi_{85}Sb_{15}$ topological insulator at complementary time scales: (1) through measurements of FMR-induced spin pumping at GHz frequencies and conversion into a dc ISHE current, and (2) femtosecond laser-induced spin injection followed by an ultrafast spin-to-charge current conversion by means of ISHE leading to THz charge transients. Our FM-TI heterostructures have been deposited using DC magnetron sputtering at room temperature. We utilize broadband FMR measurements to determine the enhancement in Gilbert damping of the precessing magnetization due to spin pumping across the FM-TI interface. The ISHE measurements confirm that this spin pumping process leads to spin-to-charge conversion in the BiSb layer, which manifests as a dc voltage ($V_{dc}$) at the resonance. From the measurements of the BiSb layer thickness dependence of $V_{dc}$ we extract the spin diffusion length ($\lambda_{SD}$) and spin-Hall angle ($\theta_{SH}$) of polycrystalline BiSb, which are 7.86 ± 1.2 nm and 0.010 ± 0.001 respectively. Measurements of THz emission when the bilayer is irradiated with femtosecond pulses of a Ti-sapphire laser further confirm spin-to-charge conversion in the BiSb layer. The BiSb layer thickness dependence of THz intensity is consistent with the measurements of $V_{dc}$.

## II. Experimental section

### A. Sample growth

Thin films with stack structure of sapphire/FeGaB(6nm)/BiSb (0, 1, 2, 3, 4, 6, 8, 10, 15 and 20 nm) were deposited by DC magnetron sputtering of $Fe_{78}Ga_{13}B_9$ (99.99%) and $Bi_{85}Sb_{15}$ (99.99%) targets in a load-lock vacuum system with a base pressure of $1x10^{-7}$ Torr. The sapphire substrates were cleaned sequentially using acetone and isopropyl alcohol in an ultrasonic bath for 15 mins, before loading into the sputtering chamber. First FeGaB films were deposited at the rate of 0.03 nm/s on (0001) sapphire at room temperature where DC power was set to 50 W. Subsequently, BiSb layers were grown on top of the FeGaB surface at the rate of 0.11 nm/s. The sputtering power for BiSb was set to a lower value (≈ 25 W) to avoid surface melting of the target. This two-step sputtering process was carried out at 5 mTorr of Ar pressure with a 5 SCCM flow rate of 99.9995% pure Ar gas. Finally, to protect the FeGaB/BiSb heterostructure from oxidation, a thin (~ 2 nm) layer of MgO was deposited by ablating a polycrystalline target of MgO with 1000 mJ pulses of a KrF excimer laser. Unless otherwise stated, the labelling A6 and A6-B1 to A6-B20 will be used for identification of the samples with thickness of BiSb ranging from 0 to 20 nm respectively. Our samples are of square and rectangular shape with lateral dimensions of 5.0 x 5.0 $mm^2$ and 0.5 x 3.0 $mm^2$, respectively.

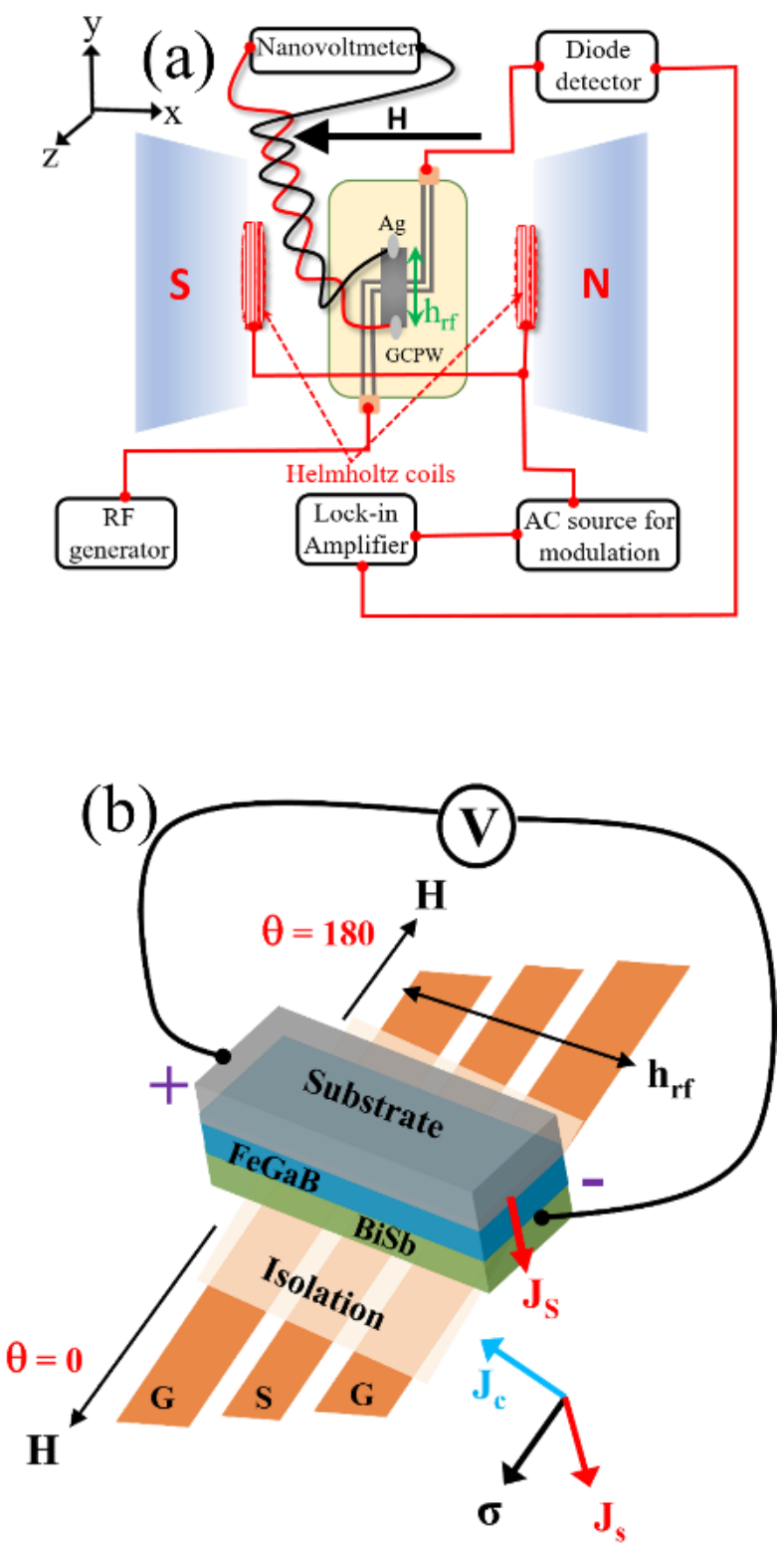

FIG. 1. (a) Schematic illustration of the experimental setup used for FMR and ISHE measurements. FMR measurements are performed over a frequency range of 10 to 30 GHz by low amplitude modulation of the dc field at 650 Hz. FeGaB/BiSb bilayer is placed in a flip-chip configuration on the signal line of the grounded coplanar waveguide. A twisted pair of copper wires is used to measure the DC voltage generated across the sample. The DC voltage is measured along y-axis with Y+ contact of the sample connected to the positive terminal of the nanovoltmeter**. (b)** Magnified view of the measurement setup to show the directions of $\mathbf{J_s}$, $\mathbf{J_c}$ and spin polarization vector (**σ**), where $\mathbf{J_c}$ = $\theta_{SH}(\mathbf{J_s}$ x **σ**). Direction of **σ** and $\mathbf{J_c}$ are controlled by changing the direction of dc magnetic field from θ = 0 to 180 deg.

### B. Experimental setup

The static magnetization (M) of the square shaped samples was measured with a vibrating sample magnetometer (Microsense Model 10 Mark II, VSM) at room temperature with the magnetic field (H) aligned parallel to the plane of the film. The combined FMR/ISHE measurements were performed with a broadband grounded coplanar waveguide (GCPW) based spectrometer over a frequency range of 10-30 GHz and RF input power ranging from 0 to +15 dBm. A schematic of our home built FMR/ISHE setup is shown in Fig. 1(a). The 0.5 x 3 $mm^2$ sample is placed in a flip chip configuration on the signal line of the GCPW. The position of the sample on GCPW is such that the external dc magnetic field is perpendicular to the microwave magnetic field component ($h_{rf}$). A 100 μm thick mica sheet was inserted between the sample surface and signal line of GCPW for electrical isolation. This precaution avoided shorting of GCPW due to the conducting nature of FeGaB and BiSb films. A small amplitude ac field generated by a pair of Helmholtz coils modulates the dc magnetic field at 650 Hz and the FMR response is recorded using a lock-in amplifier. Two thin (~40 μm diameter) copper wires are connected using silver paste at the corners of the sample [shown in Fig. 1(a)]. The generated dc voltage is measured using a nanovoltmeter. Figure 1(b) shows the cross-sectional schematic view of the FMR and ISHE measurements. The ISHE voltage was measured for two different orientations of magnetic field i.e., at θ = 0 and θ = 180 deg as shown in the figure where $\mathbf{J_s}$, $\mathbf{J_c}$ and $\boldsymbol{\sigma}$ represent the spin current, charge current, and spin polarization vectors, respectively.

To correlate the results obtained in the gigahertz frequency range, we performed THz emission measurements on the same set of samples using time-domain terahertz (THz) spectroscopy. The samples are excited by a femtosecond pulse laser with center wavelength of 800 nm and repetition rate of 10 kHz. A mechanical chopper with chopping frequency of 373 Hz is placed in the pump light path that acts as a reference for lock-in amplifier readout. The laser beam is directed perpendicular to the sample plane, with an in-plane magnetic field of ~800 Oe to align the FeGaB magnetization, as shown in the inset of Fig. 7(a). The pump light is focused on the sample which results in a fluence of ~1 mJ/$cm^2$. The emitted THz pulse and the probing laser pulse are then focused on a 1 mm thick (110) ZnTe crystal for detection by electrooptical sampling. The incident laser pulse width (full width at half maximum) is 35 fs; therefore, we measure all samples in the time-domain with a step size of 50 fs in air.

## III. Results and discussion

### A. Crystallographic structure and static magnetization characteristics of BiSb films

The crystallographic structure and electronic transport properties of the sputter deposited BiSb thin films have been reported in detail earlier [17]. The **θ-2θ** X-ray diffraction patterns show the stabilization of the rhombohedral structure (space group $R\bar{3}m$) of BiSb, with a (00l) texture. The temperature dependence of electrical resistivity and Hall coefficient of these films reveal a semi-metallic character of n-type conduction with carrier density and mobility of 1.97 x $10^{19}$ electrons/$cm^3$ and 203 $cm^2$ $V^{-1}s^{-1}$ respectively. The topological nature of its electronic states is revealed by the presence of a robust planar Hall effect which changes sign from a negative to a positive value on cooling below ~ 180 K [17].

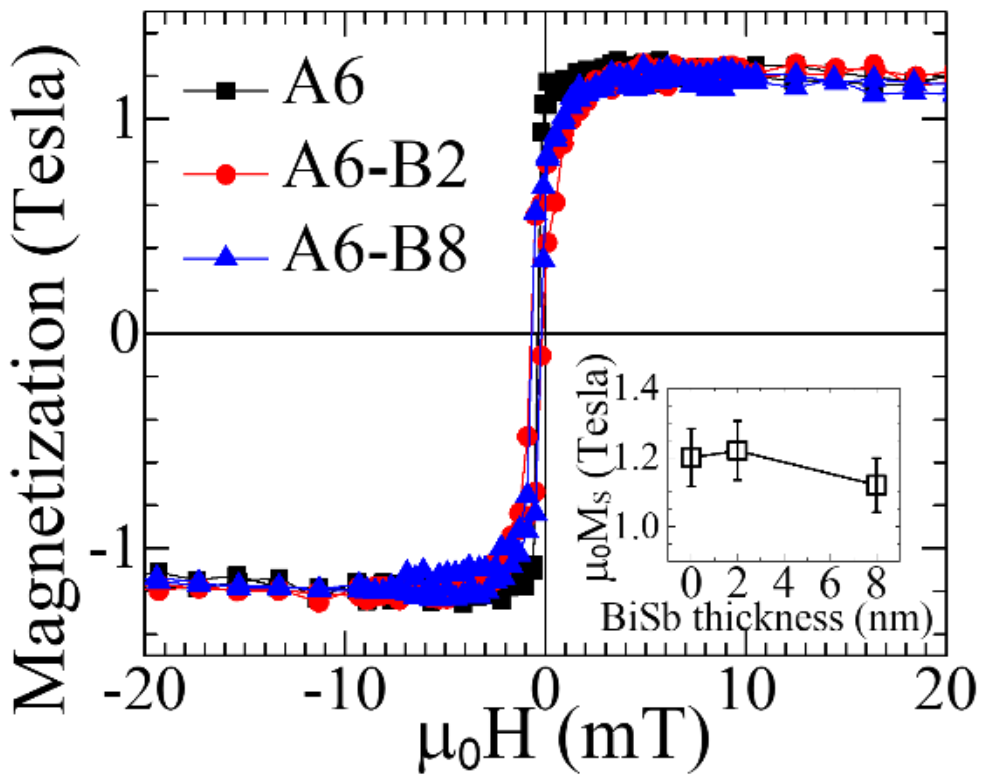

FIG. 2. The M-H loops of 6 nm thick FeGaB and FeGaB(6)/BiSb (2 nm and 8 nm) bilayers measured with external magnetic field parallel to the film plane. Inset shows the change in saturation magnetization of the bilayers as a function of BiSb thickness($t_{TI}$).

In Fig. 2 we compare the in-plane magnetization loops [M(H)] of three representative samples A6, A6-B2 and A6-B8. The magnetization in all three cases reaches a saturation value $M_s$ of 1.2 ± 0.14 T at $\mu_0 H \leq 2$ mT indicating a soft magnetic character of the FeGaB films with in-plane magnetization, which is in sharp contrast to the magnetic state of binary $Fe_xGa_{1-x}$ alloys and their thin films that display a large coercivity and high saturation magnetization [18, 19]. The addition of boron to FeGa softens its magnetic state and reduces $M_s$ [20, 21]. While the magnetic softness of the FeGaB films is like that of amorphous FeCoB, a distinguishing feature of these amorphous alloys is their large magnetostriction, which can be exploited for electrostatic tuning of their magnetization dynamics [22, 23]. The deposition of BiSb on top of FeGaB does not affect the magnetization significantly, as seen in the inset of Fig. 2, where the error bars indicate uncertainty in the measurement of the sample volume. We have also measured the magnetization of these films by rotating the in-plane magnetic field from 0 to 2π. An isotropic response was observed, which suggests the absence of any preferred axis of magnetization in the plane of the film.

## B. Dynamic magnetization probed with FMR

We now present the GHz magnetization dynamics of the FeGaB-BiSb heterostructures probed using FMR measurements over a broad frequency range of 10-30 GHz. The differential absorption (dP/dH) spectra were measured in a field sweep mode and a distinct single derivative peak was observed over the entire field range. The dP/dH curves of the FeGaB samples with different BiSb thickness collected at 10 GHz for an in-plane magnetic field are shown in the inset of Fig. 3(a). Figure.

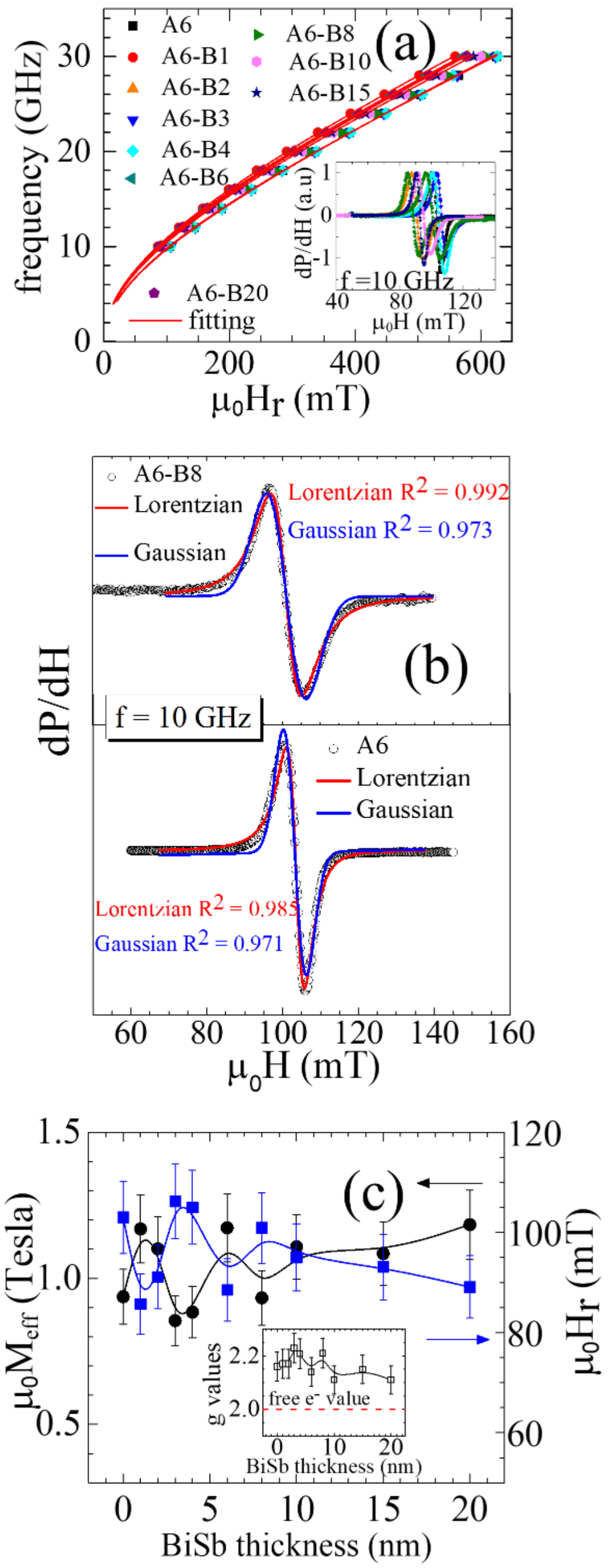

FIG. 3. (a) The FMR resonance field ($\mu_0 H_r$) plotted as a function of microwave frequency for FeGaB/BiSb bilayers of different BiSb layer thickness. Symbols represent the resonance field calculated from FMR lineshape analysis and solid lines are the fit to Kittel equation (please see text). Inset shows the differential FMR signal at 10 GHz with Lorentzian line-shape fitting. (b) FMR profiles at 10 GHz fitted with Lorentzian and Gaussian function. (c) shows the variation of effective saturation magnetization ($\mu_0 M_{eff}$) and FMR resonance field ($\mu_0 H_r$) at 10 GHz as a function of $t_{TI}$. The variation of g-factor with $t_{TI}$ is shown as inset where the dotted red line represents the free electron g-value.

3(b) shows the comparison of Lorentzian and Gaussian fitting of FMR curves at 10 GHz. The quality of fit is much better for the Lorentzian function. This is indicative of a high degree of magnetic homogeneity of the FeGaB films [24,25]. The FMR line shape at different frequencies is considered a superposition of symmetric and anti-symmetric Lorentzian functions as expressed by [26, 27],

$$\frac{dP}{dH} = K_1 \frac{4\Delta H(H-H_r)}{[4(H-H_r)^2+(\Delta H)^2]^2} - K_2 \frac{(\Delta H)^2-4(H-H_r)^2}{[4(H-H_r)^2+(\Delta H)^2]^2} \qquad (1)$$

where $K_1$ and $K_2$ are the symmetric and anti-symmetric Lorentzian coefficients. H is the applied dc field, ΔH is the FMR linewidth and $H_r$ is the resonance field. The dP/dH data at all frequencies have been fitted using Equation 1 to calculate ΔH and resonance field $H_r$. The main panel of Fig. 3(a) shows the variation of $H_r$ as a function of the frequency (f) of microwave excitation. The $H_r$ vs f data follow the Kittel equation [28, 29],

$$f = \frac{\gamma}{2\pi}\mu_0\sqrt{H(H+M_{eff})} \qquad (2)$$

where $\gamma = \frac{g\mu_B}{h}$. $g$, $\mu_B$ and $h$ are the Lande's 'g' factor, Bohr magneton and Planck's constant, respectively. Here $\mu_0 M_{eff} = \mu_0(M_S - H_S)$. We used $\mu_0 M_{eff}$ and $g$ values as free parameters while fitting the experimental data to Equation 2. The changes in $\mu_0 M_{eff}$ and $\mu_0 H_r$ (at f = 10 GHz) as a function of BiSb thickness are shown in Fig. 3(c). The absence of any significant change in $\mu_0 M_{eff}$ and $\mu_0 M_S$ with the addition of the BiSb layer suggests a metallurgically clean interface with the FeGaB layer. Figure 4(a) shows the dependence of FMR linewidth ΔH on the excitation frequency. The full width at half maximum of the resonance peak (ΔH) derives contributions from the intrinsic and extrinsic source of damping of the precessing magnetization vector. In the framework of Landau-Lifshitz-Gilbert (LLG) model for damping, the ΔH is expressed as [30],

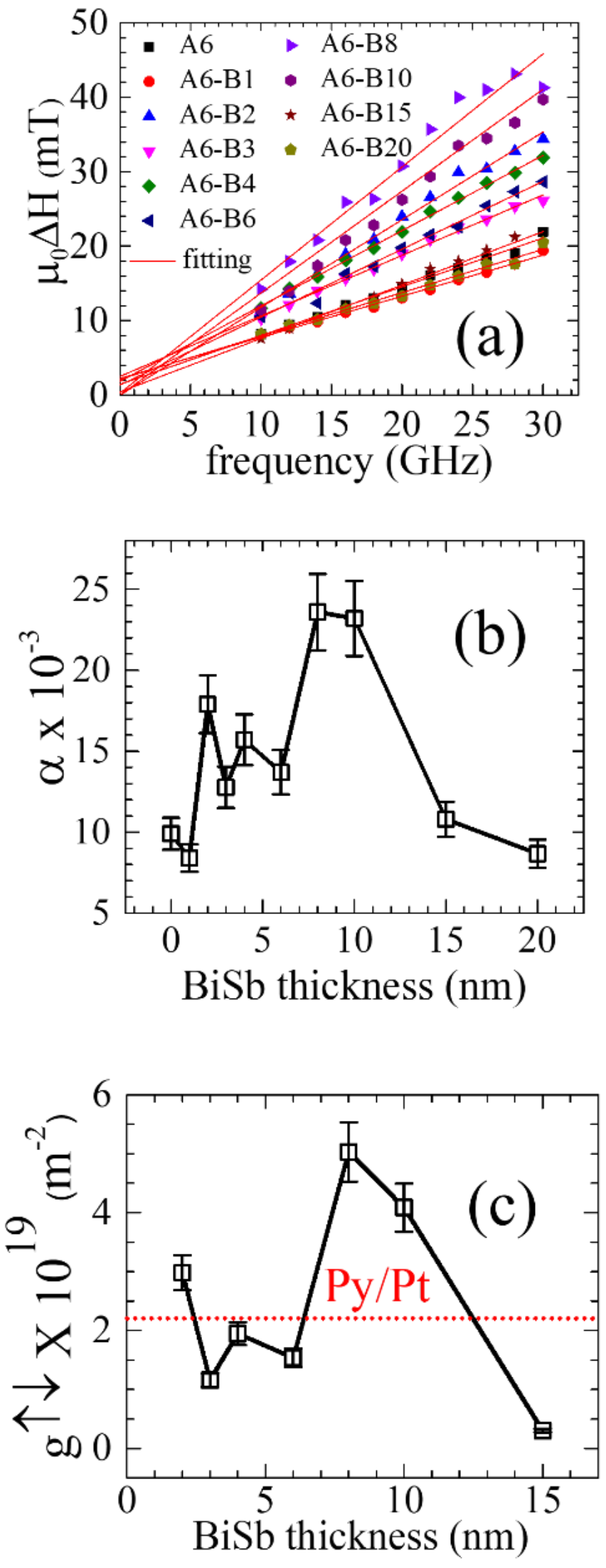

FIG. 4. (a) Change in FMR linewidth (ΔH) as a function of microwave frequency for FeGaB/BiSb multilayer of different BiSb layer thickness. Symbols are calculated ΔH while red lines are LLG fitting. (b) Variation of the Gilbert damping constant (α) as a function of $t_{TI}$ (c) Intermixing conductance ($g^{\downarrow\uparrow}$) calculated from Equation. 4 is plotted as a function of BiSb layer thickness. Dotted line in the figure is the value of $g^{\downarrow\uparrow}$ for Py/Pt bilayer [31].

$$\Delta H = \Delta H_0 + \frac{4\pi f \alpha}{\gamma} \qquad (3)$$

where, $\Delta H_0$ is the zero-field line width resulting from frozen-in defects in the material and α is the Gilbert damping constant. As seen in Fig. 4(a), the FMR linewidths are clearly enhanced in FeGaB/BiSb bilayers relative to the ΔH of the bare FeGaB film. This increase in linewidth is indicative of spin pumping that allows transfer of spin angular momentum from FeGaB to BiSb. Fitting of experimental data shown in Fig. 4(a) to Equation 3 yields the damping coefficient α, which is plotted as a function of BiSb layer thickness in Fig. 4(b). While the value of α for the bare FeGaB film agrees with previous reports [21], a strong enhancement is seen as the surface of FeGaB is covered with the topological insulator BiSb. The rapid rise of α as the BiSb layer thickness increases indicates efficient absorption of spin current generated by the precessing magnetization of FeGaB in the BiSb layer. While an enhancement in damping may also occur due to atomic interdiffusion at the interface, this possibility appears unlikely because the BiSb layers were deposited at ambient temperature. The absence of a significant metallurgical activity at the interface is also established by the static magnetization measurements as discussed earlier.

The spin pumping efficiency of an interface can be evaluated by calculating the real part of the spin mixing conductance $g^{\downarrow\uparrow}$ using the following relation [31, 32],

$$g^{\downarrow\uparrow} = \frac{4\pi M_s t_{FeGaB}}{g\mu_B}\left(\alpha_{FeGaB/BiSb} - \alpha_{FeGaB}\right) \quad (4)$$

The maximum $g^{\downarrow\uparrow}$ is calculated to be 5.03 x $10^{19}$ $m^{-2}$ for $t_{BiSb}$= 8nm. This large value of $g^{\downarrow\uparrow}$ suggests a high efficiency of spin pumping into the BiSb layer. Note that the trend in Fig. 4(b) and 4(c) is different from the behavior of normal metals like Pt [33]. In normal metals, $g^{\downarrow\uparrow}$ and α saturate with the increasing the metal layer thickness, but with BiSb these values showed the maxima up to certain thickness. Such unconventional thickness dependence of $g^{\downarrow\uparrow}$ and α suggests that topological surface states play an important role in the damping enhancement [9, 34]. The obtained $g^{\downarrow\uparrow}$ values are close to the values reported for the Py/Pt interface [31].

### C. Inverse spin Hall voltage measurements

To establish that the enhanced damping of the precessing magnetization is indeed due to spin pumping across the interface, we measure the induced dc voltage in FeGaB/BiSb bilayers in the FMR geometry at 10 GHz. The three primary sources of this dc signal are (i) ISHE in BiSb layer due to spin pumping, (ii) self-induced ISHE voltage in FeGaB due to its non-zero spin orbit coupling and (iii) rectification of rf currents because of the anisotropic magnetoresistance (AMR) and anomalous Hall effect in the FeGaB layer. However, in the geometry of our experiment the AMR results in an antisymmetric signal with respect to the resonance field, whereas the ISHE voltage is symmetric about the $H_r$ [35]. We also note that the dc signal generated in bare films of FeGaB is small with polarity opposite to that in FeGaB/BiSb bilayers. In the subsequent analysis, we ignore the small contribution of bare FeGaB and proceed by fitting the output signal to the following equation, which is a superposition of symmetric and asymmetric Lorentzian functions [36],

$$V_{mix} = K_s \frac{\delta H^2}{(H-H_r)^2+\delta H^2} + K_{as} \frac{-2\delta H(H-H_r)}{(H-H_r)^2+\delta H^2} \qquad (5)$$

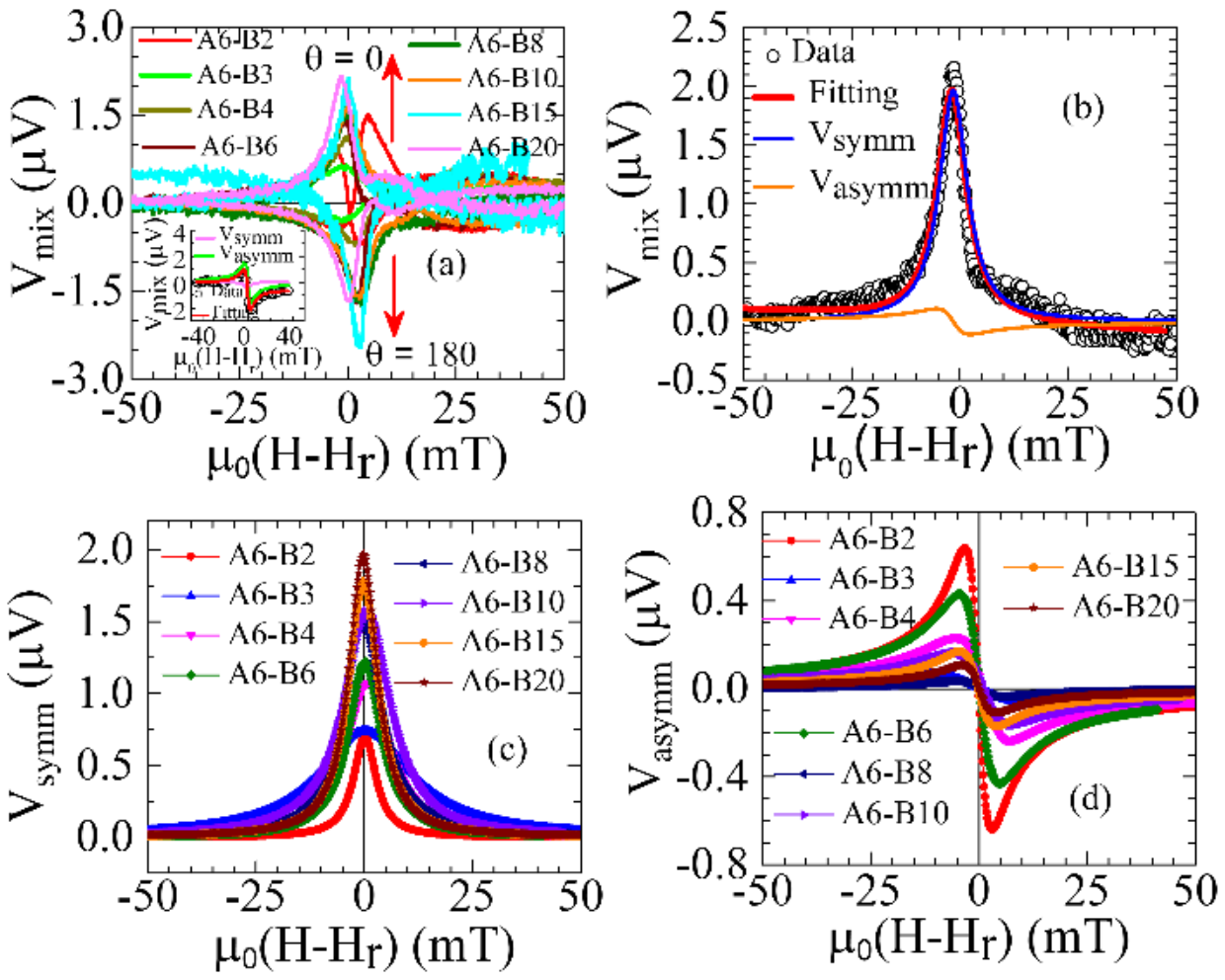

FIG. 5. (a) The ISHE related dc voltage signal at 10 GHz for different BiSb layer thickness. The $V_{dc}$ changes polarity on reorienting the field from θ = 0 to 180 deg, inset shows the signal for bare FeGaB film. (b) The $V_{dc}$ is fitted to a function made of symmetric and asymmetric Lorentzian lineshapes. Symbols represent data for the A6-B20 sample. (c) The ISHE related Symmetric component of $V_{dc}$ for bilayers of different $t_{TI,}$ and (d) AMR induced asymmetric voltage plotted for different BiSb layer thickness. The frequency of measurement for the data shown in (c) and (d) is 10 GHz.

Here $K_s$ and $K_{as}$ are symmetric and asymmetric coefficients, and δH is the half-width at half maximum; another parameter appearing in Equation 5 has been defined earlier in the context of Equation 1. The voltage signals of a series of FeGaB/BiSb samples measured at 10 GHz are shown in Fig. 5(a) for θ = 0 and 180 deg orientations of the external dc field with respect to the direction of rf current in the waveguide. The polarity of the measured voltage is governed by the relation $\boldsymbol{J_c} = \theta_{SH}(\boldsymbol{J_s} \times \boldsymbol{\sigma})$ where $\mathbf{J_c}$, $\mathbf{J_s}$ and $\boldsymbol{\sigma}$ are charge current, spin current and spin polarization vector, respectively. The polarity of $V_{dc}$ establishes that the spin Hall angle of BiSb is positive. In the inset in Fig. 5(a) we show the dc voltage output of a bare FeGaB film at 10 GHz excitation. It is important to note that the polarity of the symmetric signal in bare FeGaB is opposite to that of the FeGaB/BiSb bilayer suggesting that the actual spin-to-charge conversion in BiSb is even stronger. Our detailed analysis of self-induced ISHE in FeGaB films [37] reveals that a magnetic dead layer at the interface of film and substrate breaks the symmetry of the two surfaces and promotes a unidirectional flow of pumped spin current towards the substrate. The intrinsic SOC of FeGaB then produced a dc voltage through the ISHE. However, the presence of the topological insulator BiSb on top of the FeGaB allows a dissipation channel for the spin current pumped by the precessing magnetization towards the top surface of the magnetic layer. We note that the symmetric component of the measured signal increases as the BiSb layer is made thicker while the asymmetric AMR signal decreases [Fig. 5(c) and 5(d)]. As we will discuss below, the opposite polarity of FeGaB and FeGaB/BiSb is also confirmed in the time-domain terahertz (THz) spectroscopy (TDTS) measurements.

In the following, we estimate the spin diffusion length in BiSb by fitting the charge current ($I_c$) data deduced from the measured $V_{ISHE}$ and sample dimensions, to the following equation [33],

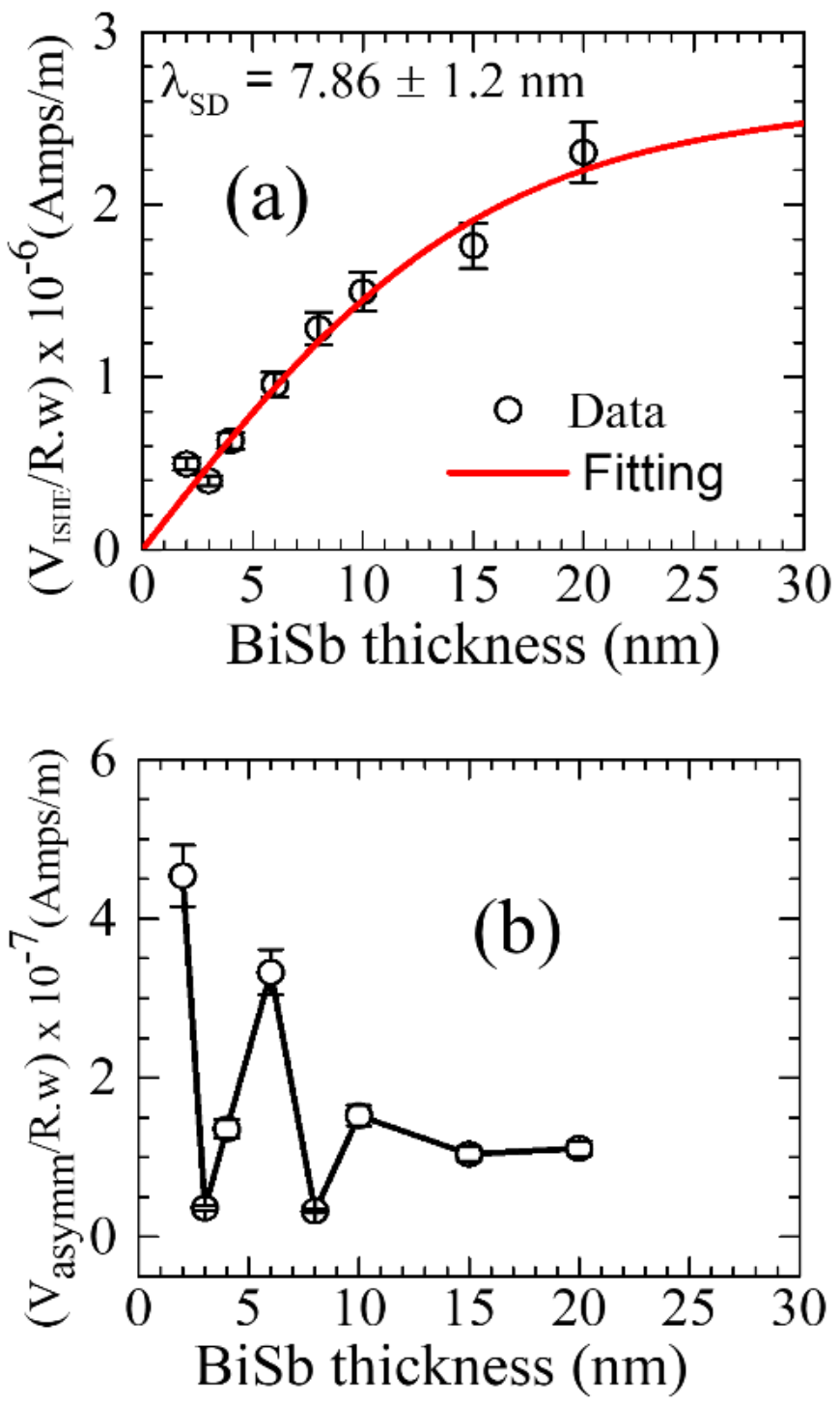

FIG. 6. (a) ISHE induced charge current normalized by sample width plotted as a function of $t_{TI}$. A fitting of these data to Equation. 6 yields the spin diffusion length in BiSb ($\lambda_{SD}$ = 7.86 ± 1.2 nm). (b) AMR and other spin rectification effects induced charge current normalized by sample width is shown for bilayers of different BiSb layer thickness.

$$\frac{I_C}{w} = \frac{V_{ISHE}}{R.w} \propto \lambda_{SD}.\tanh\left(\frac{t_{TI}}{2\lambda_{SD}}\right) \qquad (6)$$

where R, w and $t_{TI}$ are resistance, width, and thickness of the BiSb/FeGaB bilayer, and $\lambda_{SD}$ is the spin diffusion length. The result of this analysis, which yields $\lambda_{SD}$ = 7.86 ± 1.2 nm, is displayed in Fig. 6(a). Figure 6(b) shows the variation of the charge current estimated from the asymmetric component of $V_{mix}$ as a function of BiSb layer thickness. The charge current behaves like the response of a damped oscillator as $t_{TI}$ increases. This remarkable effect appears to be a result of the opposite signs of AMR in FeGaB and BiSb. The AMR in BiSb undergoes a sign change from positive to negative on increasing the temperature beyond ~ 170 K [17], whereas the AMR of FeGaB is positive at room temperature [37].

We analyze the ISHE data further to calculate the spin Hall angle of BiSb in the framework of the following equation [32],

$$V_{ISHE} = \frac{-e\theta_{SH}}{\sigma_{TI}t_{TI}+\sigma_F t_F}\lambda_{SD}\tanh\left(\frac{t_{TI}}{2\lambda_{SD}}\right) g^{\uparrow\downarrow} f L P \left(\frac{\gamma h_{rf}}{2\alpha\omega}\right)^2 \quad (7)$$

where $V_{ISHE} = V_{symm}$, e is the electron charge, $\sigma_{TI}$ ($\sigma_F$) is the conductivity of topological insulator BiSb (FeGaB), $t_{TI}(t_F)$ is the thickness of BiSb (FeGaB) layer, $g^{\downarrow\uparrow}$ is the interfacial spin mixing conductance, ω = 2πf is the FMR frequency, L is the sample length and $h_{rf}$ = 0.17 Oe is the RF magnetic field component in our GCPW-FMR setup at RF input power = +15dBm. The factor P is the ellipticity of magnetization precession in these samples calculated as [31],

$$P = \frac{2\omega[\gamma 4\pi M_S+\sqrt{(\gamma 4\pi M_S)^2+4\omega^2}]}{(\gamma 4\pi M_S)^2+4\omega^2} = 0.9 \quad (8)$$

The $\theta_{SH}$ of BiSb calculated by inputting all experimental and material parameters into Equation 7 comes out to be 0.007 ± 0.001 and 0.010 ± 0.001 for the 8 nm and 10 nm thick BiSb layers respectively. Consequently, the spin current density $J_S$ can be estimated using [32, 33],

$$J_S = \left(\frac{2e}{\hbar}\right).\frac{g^{\uparrow\downarrow}\gamma^2\hbar h_{rf}^2}{8\pi\alpha^2}\frac{[\gamma 4\pi M_S+\sqrt{(\gamma 4\pi M_S)^2+4\omega^2}]}{(\gamma 4\pi M_S)^2+4\omega^2} \quad (9)$$

The spin current density for BiSb thickness 8 nm and 10 nm was found to be 1.18 x $10^5$ A/m$^2$ and 1.01 x $10^5$ A/m$^2$ respectively. It must be noted here that previous report on BiSb showed a very large value of $\theta_{SH}$ = 52 and 22 measured in MnGa/BiSb and CoTb/BiSb multilayers [8]. Our values of $\theta_{SH}$ in BiSb is much smaller, one possibility of these discrepancies could be that previous $\theta_{SH}$ estimation of BiSb was done using second harmonic Hall measurements and dc transport measurements. These measurements involve a direct current injection through TI material, where spin momentum locking can play an important role. Deorani et al. observed a huge discrepancy in the spin Hall angle of $Bi_2Se_3$ calculated using ISHE measurements [38]. Sagasta et al. showed that spin Hall angle and spin diffusion of spin sink metal (Pt) can be largely varied with the change in conductivity [39]. We rule out this possibility because the conductivity of BiSb films obtained in this work is of the order of $10^5$ $\Omega^{-1}m^{-1}$, same as reported in Ref. [8].

### D. THz emission measurements

In addition to the microwave driven spin pumping, we study the ultrafast spin-current injection from FeGaB into BiSb of various thicknesses upon femtosecond laser pulse excitation by time-resolved THz spectroscopy.

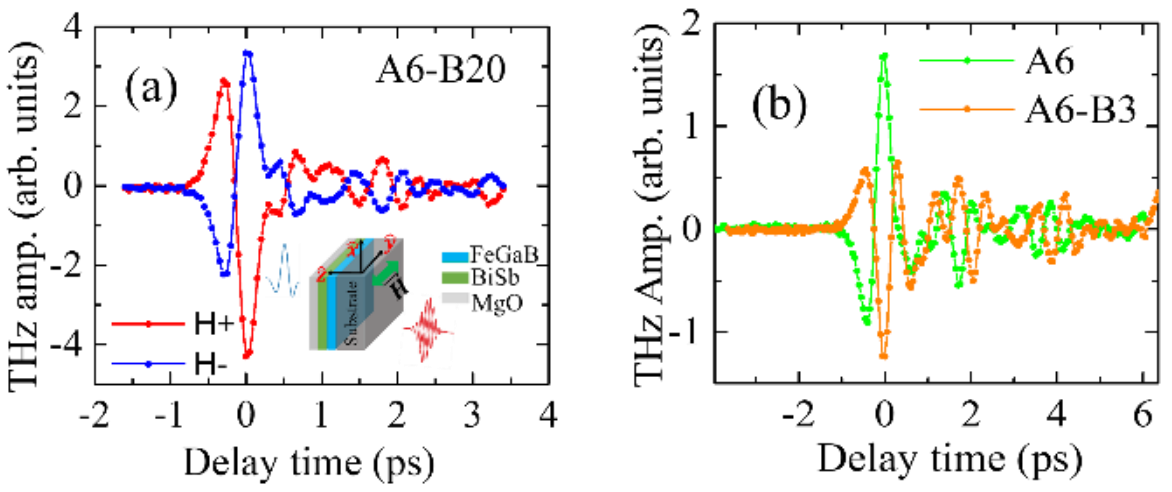

FIG. 7. (a) THz emission from FeGaB(6)/BiSb(20) bilayer film. The time traces are obtained using the time-domain THz spectroscopy system while applying in-plane magnetic fields in the opposite directions. The inset shows the experimental configuration. The laser pulse travels along the z-axis and hits vertically on the back side of the sample. The magnetic field is in-plane and parallel to the y-axis. (b) The comparison of THz emission spectra from bare FeGaB(6) layer and FeGaB(6)/FeGaB(3) bilayer. THz signal from the bilayer leads the signal from the single layer sample by around 0.3 ps. After correction, the two trance show a 180 deg phase shift.

Upon excitation with the laser pulse, an ultrafast spin current $j_S$ is launched in the ferromagnetic FeGaB layer and diffuses into the adjacent BiSb layer, in which it is converted into a charge current $j_C$ by the inverse spin

Hall effect. This charge current subsequently gives rise to THz transients according to the far-field approximation of electric dipole radiation $E_{THz}{\sim}\partial j_C/\partial t$. This THz transient is detected using standard time-domain THz spectroscopy [40]. Figure 7(a) shows the measured THz time traces from a FeGaB(6)/BiSb(20) bilayer film. The signal inverts when the magnetization direction is reversed, supporting our interpretation in terms of a rectification by inverse spin Hall effect: the spin polarization vector $\sigma$ changes sign when the FeGaB magnetization is reversed resulting in an inversion of the THz signal according to $\boldsymbol{j_C} \sim \boldsymbol{j_S} \times \boldsymbol{\sigma}$. The Fourier transform of the time-domain THz signal reveals a typical peak frequency of 0.7 THz and a 5-dB bandwidth of 1.7 THz for our samples. We compare the signal from bare FeGaB with the FeGaB/BiSb bilayer, see Fig. 7(b). Similarly, to the spin pumping/ISHE results discussed above, the signal of the bare FeGaB is inverted for the same magnetization direction, i.e. $\sigma$, illustrating that the physical mechanism of the THz generation process in the two systems is different. In the FeGaB/BiSb bilayer samples, spin-to-charge conversion occurs in the BiSb layer giving rise to THz transients by means of the ISHE. Conversely, the THz emission from bare FeGaB is explained by a contribution from three main effects: (1) magnetic dipole radiation [41, 42], (2) ultrafast demagnetization [43], and (3) ultrafast self-induced ISHE [44-46]. In the following, we present the THz emission from FeGaB/BiSb bilayers of varying BiSb thickness. In the time domain, the traces for different samples show changes in THz magnitude when the thickness of the BiSb layer is altered. However, we do not observe a phase shift or shift of time zero (the time at which the highest peak is detected). This can be explained by the fact that the direction of the injected spin current and its spin-polarization vector remain unchanged among the different sample thicknesses.

To be able to better correlate the GHz spin pumping results with the THz spin injection results discussed here, we convert the detected BiSb thickness-dependent maximum THz amplitude $E_{THz}$ in an *effective charge current* by calculating $\frac{E_{THz}}{R}$, where $R$ is the dc resistance of the bilayer samples. We note that the maximum THz amplitude in the time traces appears at an identical time in the timeline for all samples. As is seen in Fig. 8, the thickness dependence of the THz amplitude shows the same trend as the spin-pumping induced DC charge current observed in Fig. 6 suggesting that the despite the different time scales the same underlying physics are at play: spin current is pumped from FeGaB into the BiSb capping and then converted into a charge current by ISHE, where interfacial effects such as the interfacial spin-mixing conductance determine the spin-injection efficiency across the interface.

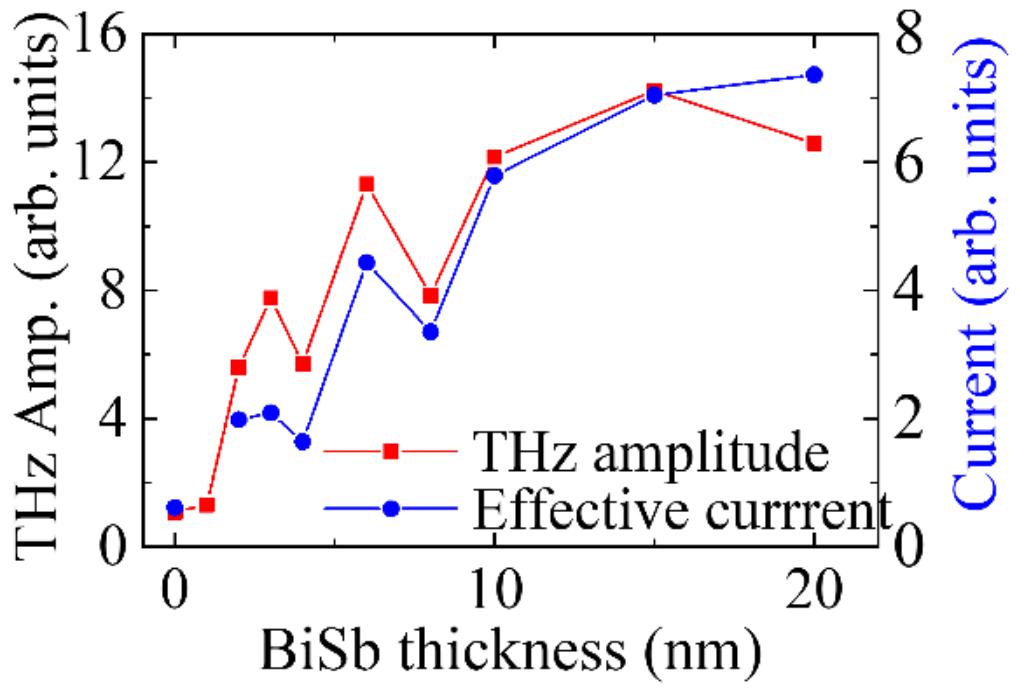

FIG. 8. The THz amplitude varies with BiSb thickness. The electric field strength at time zero from the time trace is plotted with respect to the BiSb thickness. On the same graph, the effective current (calculated by dividing the THz amplitude with the Ohm resistance of the sample) is also shown.

**E. Tight binding model of BiSb**

To shed light on our experimental results and further strengthen the correlation between GHz and THz spin pumping and subsequent spin-to-charge conversion, we performed model tight-binding calculations to investigate the thickness-dependent spin-Hall conductivity (SHC) of BiSb. We follow the model developed by Dang et al [47]. showing the proportionality of the THz signal $E_{THz}$ to the spin-Hall conductivity $\sigma_{SHE}$ of a non-magnetic material and the injected spin current $j_s$ from the ferromagnetic layer. Assuming linear response ($E_{THz} \sim \sigma_{SHE} j_s$), one then considers the SHC within the linear response theory calculated in framework of the Kubo-Bastin formula: [48, 49]

$$\sigma_{\mu\nu}^{\xi} = \frac{e^2\hbar}{2\pi\Omega}\int f(\varepsilon)Tr\big[\hat{v}_\mu(\partial_\varepsilon \hat{G}^R)\{\hat{v}_\nu,\sigma^\xi\}(\hat{G}^R-\hat{G}^A) - \hat{v}_\mu(\hat{G}^R-\hat{G}^A)\{\hat{v}_\nu,\sigma^\xi\}(\partial_\varepsilon \hat{G}^R)\big]d\varepsilon \quad (10)$$

where $\hat{G}^{R,A} = \left[\varepsilon - \hat{H} \pm i\Gamma\right]^{-1}$is the retarded (R) and advanced (A) Green's function, respectively, with energy $\varepsilon$,$\Gamma$ the homogeneous broadening, $H$ is the Hamiltonian of the system, $Tr$ indicates the trace over the wavevector **k** and the band index, e is the electron charge, $\Omega$ is the volume of the nonmagnetic material, $\hat{v}_{\mu\nu} = (1/\hbar)\partial_{k_{\mu\nu}}\hat{H}$ is the velocity operator, $\sigma^\xi$ the Pauli matrices $(\mu,\nu,\xi \equiv x,y,z)$, and $f(\varepsilon)$ is the Fermi distribution function.

We employed the $sp^3$ three nearest neighbors tight-binding (TB) model for bulk Bi and Sb crystal from Ref. [50] adapted to BiSb slabs [51]. Note that both bismuth and antimony possess a rhombohedral structure with two atoms per unit cell, and can also be represented by a hexagonal conventional unit cell containing 6 atoms, forming bilayers perpendicular to the c axis. A $Bi_{1-x}Sb_x$ alloy is obtained by replacing x percent of Bi with Sb atoms, and a BiSb thin film is created by stacking the bilayers along [0001] direction of the hexagonal unit cell. The total tight-binding Hamiltonian for N bilayers of BiSb is given in the following form:[51]

$$H = \begin{bmatrix} H_{1,1} & H_1 & & & & \\ H_1^\dagger & H_{2,2} & H_2^\dagger & & & \\ & H_1 & H_{3,3} & H_1 & & \\ & & H_1^\dagger & H_{4,4} & \ddots & \\ & & & \ddots & \ddots & H_1 \\ & & & & H_1^\dagger & H_{2N,2N} \end{bmatrix}$$

where $H_{i,i}(i = 1{:}2N)$ is the Hamiltonian of one layer (one atomic plane), $H_1$ and $H_2$ are the hopping terms between two layers decomposed into inter- and intra-bilayer hopping terms. The intra-bilayer hopping term $H_1$ contains the third and second nearest neighbor hopping terms whereas the inter-bilayer hopping term $H_2$ involves the nearest-neighbor hopping terms. The Rashba effect in the two-dimensional electron gas (2DEG) system at the pristine BiSb's surfaces is described via additional hopping terms between the

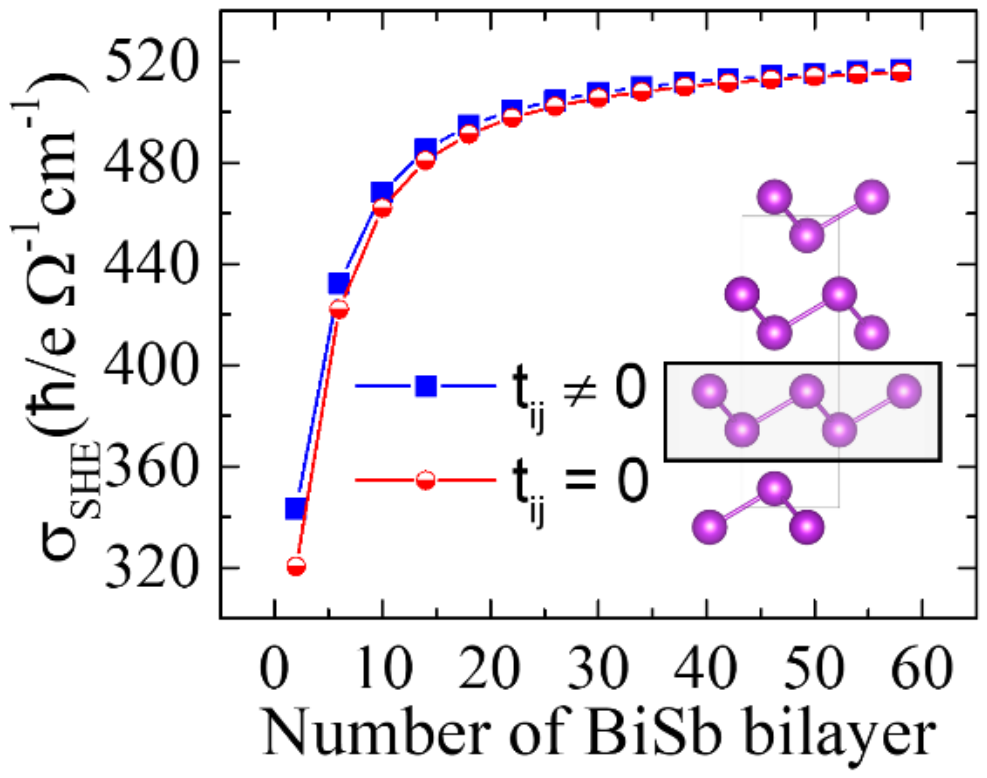

FIG. 9. The spin Hall conductivity of $Bi_{85}Sb_{15}$ vs. number of $Bi_{85}Sb_{15}$ bilayer computed with the theoretical model in section E for two cases: with and without surface Rashba effect corresponding to hopping term $t_{ij} \neq 0$ and $t_{ij} = 0$ respectively. The inset presents the crystal structure of Bi and the $Bi_{85}Sb_{15}$ bilayer structure is effectively defined by replacing 15% Bi by Sb atoms within each Bi bilayer.

nearest-neighbor sites $R_m$ and $R_n$ given by: [51, 52]

$$t_{ij}^{(\pm)} = \begin{cases} \pm\gamma_{pp}\cos\theta_{mn} & (i,j) \equiv (p_x, p_z) \\ \pm\gamma_{pp}\sin\theta_{mn} & (i,j) \equiv (p_y, p_z) \\ \pm\gamma_{sp} & (i,j) \equiv (s, p_z) \end{cases}$$

which are added to the uppermost $H_{1,1}$ and lowermost $H_{2N,2N}$ atomic layer with the sign $\pm$ respectively. Here, $\theta_{mn}$ is the azimuth angle between $R_m$ and $R_n$, $\gamma_{pp}$ and $\gamma_{sp}$ are hopping matrix elements between p-p and s-p orbitals.

The hopping parameters $V_C^{BiSb}$ for the TB model of $Bi_{1-x}Sb_x$ alloy are obtained by using the Virtual Crystal Approximation (VCA):[53]

$$V_C^{BiSb}(x) = xV_C^{Sb} + (1 - x^2)V_C^{Bi}$$

where x is the concentration of antimony; $V_C^{Sb}$ and $V_C^{Bi}$ are the respective hopping parameters of Sb and Bi taken from Ref. [50]

The calculated SHC as a function of the number of BiSb bilayers is shown in Fig. 9. Upon increasing the number of layers, the SHC rapidly rises up from ~ 340 $(\hbar/e\,\Omega^{-1}cm^{-1})$ calculated for 2 bilayers to ~ 480 $(\hbar/e\,\Omega^{-1}cm^{-1})$ for 15 bilayers, corresponding to 5 nm thickness of BiSb. Then a saturation level is found when the SHC reaches the bulk's value of about 520 $(\hbar/e\,\Omega^{-1}cm^{-1})$ for the $Bi_{85}Sb_{15}$ alloy, i.e., when the number of layers becomes large enough to represent a bulk BiSb. This thickness-dependent behavior of SHC is attributed to the fact that more channels are opened up and contribute to transport within the BiSb layer upon increasing its thickness. Furthermore, our calculations of SHC predict a minor contribution of the Rashba effect (spin-momentum locking) at the interface of pristine BiSb to its spin-Hall conductivity. Overall, the calculated thickness-dependent SHC is in good agreement with the experimental findings; both the GHz spin pumping and the THz spin injection results, i.e., as the BiSb thickness is increased, the SHC increases, which in turn leads to the larger experimentally detected inverse spin-Hall signal. We note that our calculations do not consider the interface properties between the FeGaB and BiSb. We attribute the thickness-dependent oscillations observed both in GHz spin pumping and THz spin injection to the non-monotonic behavior of the interfacial spin-mixing conductance previously reported in Ref. [54]. Indeed, Machon et al. theoretically showed that quantum oscillations [54] caused by surface states lead to such a non-monotonic thickness dependence of the spin-mixing conductance. While our calculations do not consider this effect, they explain the general trend of our experimental observations carried out in the GHz and THz frequency range.

## IV. Summary

We have successfully fabricated bilayers of polycrystalline BiSb and amorphous FeGaB at room temperature using DC magnetron sputtering. The FeGaB films display a narrow FMR linewidth and in-plane magnetization which allow pumping of a large spin current across the interface at resonance. The static dc magnetization measurements on bilayers of different BiSb layer thickness indicate absence of any interdiffusion at the interface which would affect the magnetization of the FeGaB layer. The enhanced damping of precessing magnetization in FeGaB/BiSb suggests dissipation of the pumped spin current in the BiSb attributed to an enhanced spin-orbit coupling. We systematically studied the dc voltage induced by the inverse spin-Hall effect across the length of the bilayer when it is driven to resonance. A careful analysis of the shape of $V_{dc}$ and its symmetry on field reversal allows us to calculate the contributions of AMR and ISHE to the dc voltage. From the BiSb layer thickness dependence of the ISHE signal we have calculated the spin Hall angle and spin diffusion length in BiSb, which are 0.010 ± 0.001 and 7.86 ± 1.2 nm, respectively. As a complementary experimental technique, we employed time-domain THz spectroscopy and showed that the described spin injection and spin-to-charge conversion extends to ultrafast time scales. Since FeGaB shows substantial magneto-elastic effects, this may open new opportunities for tailored THz emission based on spintronic THz emitters. The theoretical calculations of the spin-Hall conductivity based on a linear response theory in the framework of the Kubo-Bastin formula and considering an atomistic tight-binding model agrees remarkably well with the experimental results. The results presented here provide a promising platform for creating novel spin-orbit torque devices operating in a wide frequency range from GHz - THz.

**Acknowledgements**

The research at Morgan State has been supported by the Air Force Office of Scientific Research through Grant No. FA9550-19-1-0082. Research at Delaware including ultrafast spectroscopy and theoretical calculations was supported by NSF through the University of Delaware Materials Research Science and Engineering Center DMR-2011824. M.B.J. acknowledges additional support from the NSF through Grant No. 1833000. Product names are mentioned to provide an accurate record of what was done. Reference of product names does not constitute validation or endorsement.

The authors contributed in the following ways- V.S: deposition of the FeGaB and BiSb thin films, magnetization dynamics and inverse spin Hall measurements of FeGaB/BiSb multilayers; W.W: Terahertz measurements of FeGaB/BiSb multilayers; P.B: DC magnetic measurements of FeGaB/BiSb multilayers; D.Q: theoretical calculation of spin hall conductivity of BiSb thin films using tight binding model; A.J: supported the designing of FMR and ISHE experimental setup; A.Jan: supervision on theoretical tight binding calculations; G.W.B: scientific supervision and support of the work; L.G: scientific supervision and support of the work; M.B.J: supervision on THz measurements; R.C.B: planning and formulation of the scientific concept, supervision on FMR and ISHE measurements. The paper was co-written by V.S, R.C.B and M.B.J with input from the rest of the authors.

The authors declare no competing interest.